\documentclass[aps,prd,10pt,twocolumn,superscriptaddress, nofootinbib]{revtex4}
\usepackage[dvipsnames]{xcolor}
\usepackage{physics}
\usepackage{braket}
\usepackage{scalerel}
\usepackage{blindtext} 
\usepackage{epsfig, cancel}

\usepackage{latexsym}
\usepackage{natbib, comment}
\usepackage{mathrsfs,amsmath,amssymb,amsthm,amsfonts,tikz,graphicx,accents,hyperref,color}
\usepackage{url}
\usepackage{dcolumn}
\usepackage{multirow}
\usepackage{color}
\usepackage{cancel}
\usepackage{soul}
\usepackage[normalem]{ulem}
\usepackage{txfonts}
\usepackage{epsfig}
\usepackage{psfrag}
\usepackage{subfigure}
\hypersetup{colorlinks=true}
\usepackage{mathtools}
\usepackage{enumitem}
\usepackage{float}
\usepackage{caption,ragged2e}
\usetikzlibrary{decorations.markings}

\def\H0{{\text{H}\hspace*{-2.05mm}\text{H} 0\hspace*{-1.35mm}0\ }}

\usepackage[all]{xy}

\usepackage{slashed}
\usepackage{slashed,ccaption}
\usepackage{multirow}
\usetikzlibrary{calc} 

\usepackage{caption}

\usepackage{array}
\newcommand{\Ottbar}{{\textrm{T}\hspace*{-1mm}{\textrm{T}}}}

\newcommand{\cttbar}{\mathcal{T}\hspace*{-1mm}{\mathcal{T}}}

\hypersetup{ linktoc=all,
    colorlinks, linkcolor={palatinateblue},
    citecolor={red}, urlcolor={amaranth} 
}

\graphicspath{{Images/}}

\renewcommand{\d}[1]{\ensuremath{\operatorname{d}\!{#1}}}

\DeclareSymbolFont{extraup}{U}{zavm}{m}{n}
\DeclareMathSymbol{\varheart}{\mathalpha}{extraup}{86}
\DeclareMathSymbol{\vardiamond}{\mathalpha}{extraup}{87}
\makeatletter
\renewcommand*{\@fnsymbol}[1]{\ensuremath{\ifcase#1\or \clubsuit \or \vardiamond \or \varheart\or
    \spadesuit\or \mathparagraph\or \|\or **\or \dagger\dagger
    \or \ddagger\ddagger \else\@ctrerr\fi}}
\makeatother

\definecolor{rosy}{RGB}{230,235,252}
\definecolor{myframetitle}{RGB}{90,89,170}
\definecolor{myblocktitle}{RGB}{140,185,249}
\definecolor{mytitle}{RGB}{10,80,26}

\definecolor{darkgreen}{RGB}{27,130,45}
\definecolor{darkblue}{rgb}{0,0,0.3}
\definecolor{darkred}{rgb}{0.7,0,0}

\definecolor{light gray}{RGB}{220,220,220}
\definecolor{dark purple}{RGB}{108,0,217}
\definecolor{pink}{RGB}{190,20,100}
\definecolor{orang}{RGB}{193,63,0}
\definecolor{green}{RGB}{11,98,17}
\definecolor{darkpink}{RGB}{153,0,76}
\definecolor{bluegreen}{RGB}{0,102,102}
\definecolor{greenlagan}{RGB}{0,102,0}
\definecolor{redgreen}{RGB}{102,102,0}
\definecolor{Redgreen}{RGB}{153,76,0}
\definecolor{vividviolet}{rgb}{0.62, 0.0, 1.0}
\definecolor{amaranth}{rgb}{0.9, 0.17, 0.31}
\definecolor{palatinateblue}{rgb}{0.15, 0.23, 0.89}
\definecolor{brightpink}{rgb}{1.0, 0.0, 0.5}
\definecolor{cornflowerblue}{rgb}{0.39, 0.58, 0.93}
\definecolor{deepcarminepink}{rgb}{0.94, 0.19, 0.22}
\definecolor{radicalred}{rgb}{1.0, 0.21, 0.37}

\newcommand\ignore[1]{}
\usepackage[most]{tcolorbox}

\tcbset{highlight math style={left=02mm,right=02mm,top=02mm,bottom=02mm}} 
\usepackage{empheq}

\newcommand\inbox[1]{\tcbset{fonttitle=\scriptsize} \tcboxmath[colback=white,colframe=black!70]{#1}}

\begin{document}

\newcommand{\mytitle}{\begin{center}{\large{\textbf{Gravity Is Induced By Renormalization Group Flow}}}
\end{center}}

\title{{\mytitle}}
\author{H.~Adami}\email{hadami@simis.cn}
\affiliation{Center for Mathematics and Interdisciplinary Sciences, Fudan University, Shanghai,
200433, China}
\affiliation{Shanghai Institute for Mathematics and Interdisciplinary Sciences (SIMIS), Shanghai, 200433, China}
\author{M.~M.~Sheikh-Jabbari}\email{jabbari@theory.ipm.ac.ir}
\affiliation{School of Physics, Institute for Research in Fundamental Sciences (IPM), P.O.Box 19395-5531, Tehran, Iran}
\affiliation{Beijing Institute of Mathematical Sciences and Applications (BIMSA), Huairou District,
Beijing 101408, P. R. China}
\author{V.~Taghiloo}\email{v.taghiloo@iasbs.ac.ir}
\affiliation{School of Physics, Institute for Research in Fundamental Sciences (IPM), P.O.Box 19395-5531, Tehran, Iran}
\affiliation{Department of Physics, Institute for Advanced Studies in Basic Sciences (IASBS),
P.O. Box 45137-66731, Zanjan, Iran}
\begin{abstract}
We revisit the holographic renormalization group (RG) setting in which a 4-dimensional ($4d$) quantum field theory at a finite cutoff corresponds to/is described by the Einstein gravity on a part of AdS$_{5}$ space, cutoff at a finite radius.  This holographic setting has interesting and important implications for the $4d$ field theory: Deformation of the field theory by a certain combination involving the square of its energy-momentum tensor can be alternatively viewed as formulating the field theory on a background with a dynamical metric. Explicitly, starting with a non-gravitating $4d$ field theory in the UV, flowing to the IR, quantum effects that we compute using the classical $5d$ Einstein gravity theory, induce an effective $4d$ Einstein gravity theory. In other words, we show that \emph{gravity is not a fundamental force and is an effective description of quantum effects in the IR limit.} 
\end{abstract}
\maketitle
\section{Introduction}

Understanding and formulating gravity has always been one of the main questions of theoretical physics. The first formulation was Newton's seminal inverse-square law, followed by Einstein's General Relativity (GR). The salient feature of  Einstein's GR is its universality: Anything or system takes part in gravitational interactions by its mere existence, i.e, having a non-vanishing energy momentum tensor (EMT). In Einstein's GR, this universality is achieved by associating gravity with the fabric of spacetime, its geometry, and in particular, its metric, within which all physical systems reside.  

Decoupling of scales is a fundamental principle of physics, best formulated through Wilsonian formulation of quantum field theory (QFT) and the renormalization group (RG) \cite{Wilson:1973jj, Weinberg:1996kr}: Any local QFT can be defined up to a desired precision below a given energy scale (cutoff) $\Lambda$ by specifying the value of its parameters (couplings) at $\Lambda$. The key notion here is the coarse graining: we can describe the theory at lower energy scales (longer distances) by  ``integrating out'' higher energy degrees of freedom. As a result of the coarse graining, the value of the couplings of the theory at energies below $\Lambda$ is determined in terms of the values at $\Lambda$, by the renormalization group flow (RGF) equations that capture quantum (loop) effects. Besides running of the couplings, we need to deform the theory by higher dimension (irrelevant) operators. 

The other two remarkable universal frameworks, thermodynamics and hydrodynamics, can also be understood through the lens of the decoupling of scales: Any many-body system or any QFT, at sufficiently low energy and long distances (formally when a gradient expansion is applicable), admits a hydrodynamical description. 
The idea that the two very universal frameworks, Einstein's GR and hydrodynamics, should be somehow related to each other has been on the minds of many physicists in the last half a century \cite{Bardeen:1973gs, Damour:1979wya, PhysRevD.18.3598, Bhattacharyya:2007vjd, Verlinde:2010hp, Grumiller:2022qhx}. This work may be viewed as a formulation of similar ideas within the Wilsonian effective field theory (EFT) and RGF equations.  

To this end, we use the holographic renormalization \cite{deBoer:1999tgo, deHaro:2000vlm, Bianchi:2001kw, Akhmedov:2010mz}, i.e., the RG formulated within the AdS/CFT duality \cite{Maldacena:1997re}, also known as holography \cite{Witten:1998qj, Aharony:1999ti}. AdS/CFT and its most commonly used limit, the gauge/gravity correspondence, provides a framework to formulate $d$-dimensional QFT in terms of a gravity theory on $(d+1)$-dimensional anti-de Sitter  AdS$_{d+1}$ spacetime. In this correspondence, the boundary QFT is encoded in a classical gravity theory in the bulk, with the radial (or holographic) direction in AdS playing the role of the RG scale \cite{Skenderis:2002wp}. Radial evolution in the bulk then corresponds to {RGF} in the boundary theory \cite{Sheikh-Jabbari:2013AdSCFT}. While the arguments and results are by and large generic, here we focus on the $d=4$ case,  a $4d$ QFT dual to Einstein gravity on AdS$_5$. 

Our main result is that the RGF equations for the EMT of a QFT—which corresponds to a subset of the bulk Einstein equations—can be reinterpreted as the dynamical equations of a gravitational theory on the boundary. In this picture, a $4d$ QFT that is non-gravitating in the UV gives rise to an effective classical gravitational theory in the IR. More precisely, the emergence of gravity is a reformulation of the RGF driven by deformations involving the square of the EMT. Since every QFT possesses an EMT, this mechanism applies universally to any $4d$ QFT (with a presumed AdS$_5$ dual). In this sense, gravity is not fundamental but emerges as a low-energy, IR manifestation of quantum field theoretic dynamics.

\section{Einstein \texorpdfstring{AdS$_5$}{AdS5} Gravity}
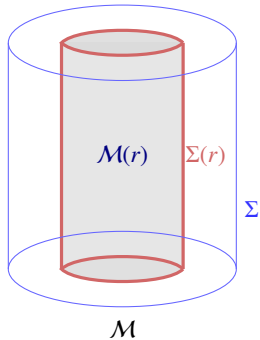
\begin{figure}[t]
\centering
\begin{tikzpicture}[scale=1.0]

\node (v1) at (0.8,0) {};
\node (v4) at (0.8,-3) {};
\node (v5) at (-0.8,0) {};
\node (v8) at (-0.8,-3) {};

\begin{scope}
  \clip ($(v1)+(0,0)$) to[out=135,in=45, looseness=0.5] ($(v5)+(0,0)$)
        -- ($(v8)+(0,0)$) to[out=45,in=135, looseness=0.5] ($(v4)+(0,0)$)
        -- cycle;
  \fill[gray!20] (-0.8,0) -- (-0.8,-3) -- (0.8,-3) -- (0.8,0) -- cycle;
\end{scope}

\draw [darkred!60, very thick] (-0.8,0) -- (-0.8,-3);
\draw [darkred!60, very thick] (0.8,0) -- (0.8,-3);

\begin{scope}[fill opacity=0.8, very thick, darkred!60]
  \filldraw [fill=gray!30] (0.8,0) 
    to[out=135,in=45, looseness=0.5] (-0.8,0)
    to[out=315,in=225,looseness=0.5] (0.8,0);
\end{scope}
\begin{scope}[fill opacity=0.8, very thick, darkred!60]
  \filldraw [fill=gray!30] (0.8,-3) 
    to[out=135,in=45, looseness=0.5] (-0.8,-3)
    to[out=315,in=225,looseness=0.5] (0.8,-3);
\end{scope}

\draw [blue!60](0,0) ellipse (1.5 and 0.5);
\draw [blue!60] (0,-3) ellipse (1.5 and 0.5);
\draw [blue!60] (-1.5,0) -- (-1.5,-3);
\draw [blue!60] (1.5,0) -- (1.5,-3);

\fill (0,-1.5)  node [blue!50!black] {${\cal M}(r)$};
\fill (0.8,-2)+(0.3,0.5) node [darkred!60] {$\Sigma(r)$};
\fill (0.8,-3)+(0.7,0.8) node [right,blue!80] {$\Sigma$};
\fill (0, -3.8) node [black] { $\mathcal{M}$};
\end{tikzpicture}
\caption{ \justifying{\footnotesize{ Portion of an asymptotically AdS spacetime: the shaded region $\textcolor{blue!50!black}{\mathcal{M}(r)}$ is enclosed by a timelike surface $\textcolor{darkred!60}{\Sigma(r)}$.  $\textcolor{black}{\cal{M}}$ denotes the global asymptotically AdS spacetime, with its asymptotic timelike boundary labeled $\textcolor{blue!80}{\Sigma}$.}}}
\label{fig:ADS-timelike}
\end{figure}

\paragraph*{\underline{Geometric Setup.}}
Let $({\cal M},g)$ denote  an asymptotically AdS$_5$ geometry, with asymptotic boundary $\partial \mathcal{M}=\Sigma$. To analyze the bulk geometry, we introduce a foliation of $\mathcal{M}$ by a family of timelike, codimension-1 hypersurfaces $\Sigma(r)$, labeled by a radial coordinate $r$ taken to be one of the spacetime coordinates, with $r \in (r_\circ, \infty)$ where $r_\circ$ stands for possible inner boundary. Each hypersurface $\Sigma(r)$ is foliated by intrinsic coordinates $x^a$, such that the relevant region of spacetime is covered by the coordinate system $\{r, x^a\}$.  $\mathcal{M}(r)$ denotes the region enclosed by a hypersurface at constant $r$. The complete spacetime is recovered in the  $r \to \infty$ limit, where $\mathcal{M}(r \to \infty) = \mathcal{M}$ and the foliation hypersurface approaches the asymptotic boundary: $\Sigma(r \to \infty) = \Sigma$; see Fig.\ref{fig:ADS-timelike}.

In the Fefferman--Graham gauge \cite{Fefferman1985, fefferman2012ambient}, the bulk metric takes the form
\begin{equation} \label{metric}
    \mathrm{d}s^2 = g_{\mu\nu} \d{}x^\mu \d{}x^\nu= \frac{L^2}{r^2} \, \mathrm{d}r^2 + h_{ab}(r, x) \, \mathrm{d}x^a \mathrm{d}x^b \, ,
\end{equation}
where $L$ is the AdS$_5$ radius, and $h_{ab}$ denotes the induced metric on the constant-$r$ slice $\Sigma(r)$. The above defines an asymptotically AdS geometry when 
\begin{equation}\label{rescale-1}
    h_{ab}(r,x^a) := \frac{r^{2}}{L^2} \gamma_{ab}(r,x^a)\, ,\qquad  \gamma_{ab}\Big|_{\Sigma} \sim \mathcal{O}(1) \, ,
\end{equation}
and $\gamma_{ab}(r\to\infty, x^a)$ is the  metric at the conformal boundary. 
The extrinsic curvature of the hypersurfaces $\Sigma(r)$, $K_{ab}$ and its trace $K$ are defined as
\begin{equation} \label{def-K}
    K_{ab} := \frac{r}{2L} \, \partial_r h_{ab} \, ,\qquad K:=h^{ab} K_{ab}= \frac{r}{L}\frac{\partial_r \sqrt{-h}}{\sqrt{-h}}\, ,
\end{equation}
where $h:=\det(h_{ab})$.

\paragraph*{\underline{Action.}}
We consider the $5d$ Einstein--Hilbert action  with a negative cosmological constant, defined over a finite radial domain $\mathcal{M}(r)$, supplemented by boundary terms consistent with Dirichlet boundary conditions (bc's):
\begin{equation}\label{action-AdS-NR}
    S\!_r[g] 
    = \frac{1}{2\,\kappa_5} \int_{\mathcal{M}(r)} \sqrt{-g} \left(R[g] + \frac{12}{L^2}\right) 
    + \frac{1}{\kappa_5} \int_{\Sigma(r)} \sqrt{-h}\, K\, ,
\end{equation}
where $\kappa_5$ denotes the $5d$ gravitational coupling constant, 
$R[g]$ is the Ricci scalar associated with the bulk metric $g_{\mu\nu}$, and the second term, the Gibbons--Hawking--York (GHY) boundary term \cite{PhysRevLett.28.1082, PhysRevD.15.2752}, is added to ensure a well-posed variational principle with Dirichlet bc on $\Sigma(r)$.

On-shell variation of the action \eqref{action-AdS-NR}  yields the boundary term
\begin{equation}\label{on-shell-var-action-D}
    \delta S\!_r[g]\Big|_{\text{on-shell}} 
    = -\frac{1}{2} \int_{\Sigma(r)} \sqrt{-h}\, T^{ab}\, \delta h_{ab}\, ,
\end{equation}
where 
\begin{equation}\label{BY-EMT}
    T^{ab} := \frac{1}{\kappa_5}\left(K^{ab} - K h^{ab}\right)\, ,
\end{equation}
is the Brown--York energy--momentum tensor (BY-EMT) \cite{Brown:1992br}. $\sqrt{-h}T^{ab}$ is the canonical conjugate to the  metric on $\Sigma(r)$, $h_{ab}$ \cite{Parvizi:2025shq}. By design, \eqref{on-shell-var-action-D} vanishes for the Dirichlet bc $\delta h_{ab}=0$.

The bulk equations of motion obtained from \eqref{action-AdS-NR} are the standard Einstein equations:
\begin{equation}\label{eom-AdS}
    R_{\mu\nu}[g] - \frac{1}{2} R[g]\, g_{\mu\nu} - \frac{6}{L^2} g_{\mu\nu} = 0\, .
\end{equation}
The $1+4$ decomposition of the Einstein equations with respect to the foliation by hypersurfaces $\Sigma(r)$ yields \cite{Parvizi:2025wsg}
\begin{subequations}\label{EoM-GR-mat-decompose-NR}
    \begin{align}
       & R[h] + \frac{12}{L^2} + \kappa_5^{2} {\Ottbar} = 0 \,, \label{EoM-ss-NR} \\
       & \nabla_b   {{T}}^{ab} = 0 \, , \label{EoM-sa-NR} \\
       & r\partial_r   {  {T}}_{ab} =
        \kappa_5\, L\left(2   {  {T}}_{a}^{\;c}   {  {T}}_{bc} 
       - \frac{1}{3}   {  {T}}   {  {T}}_{ab} 
       +   {\Ottbar} h_{ab} \right) +\frac{L}{\kappa_5}  R_{ab}[h] , \label{EoM-ab-NR}\\
       & r\partial_{r} h_{ab} = 2\,L\,\kappa_{5}\Big(   {  {T}}_{ab} - \frac{  {  {T}}}{3} h_{ab}\Big)\, , \label{Tab-def}
    \end{align}
\end{subequations}
where the last line is just definition of ${T}_{ab}$. $R[h]$ and $R_{ab}[h]$ denote the Ricci scalar and Ricci tensor of the induced metric $h_{ab}$, and $\nabla_a$ is the covariant derivative compatible with $h_{ab}$. The quantity $  {\Ottbar}$ is the generalization of the T$\bar{\text{T}}$ operator to four dimensions \cite{Taylor:2018xcy, Hartman:2018tkw, Parvizi:2025wsg}, defined as
\begin{equation}\label{TTbar-NR}
      {\Ottbar} =   {  {T}}^{ab}   {  {T}}_{ab} - \frac{1}{3}   {  {T}}^2 \, .
\end{equation}
Eqs.~\eqref{EoM-ss-NR} and \eqref{EoM-sa-NR} represent the radial Hamiltonian and momentum constraints \cite{Arnowitt:1962hi}, as they involve no radial derivatives, and \eqref{EoM-ab-NR} determines the radial flow of the BY-EMT along the holographic direction. 

\section{Renormalization Group Induced Gravity}
Having set up the stage, we begin by formulating holography at a finite radial distance. It is well known that radial evolution in the bulk of an asymptotically AdS spacetime corresponds, holographically, to the RGF in the boundary theory \cite{deBoer:1999tgo, Skenderis:2002wp}. Therefore, developing finite-cutoff holography requires moving radially inward in the bulk while simultaneously triggering RGF in the boundary theory.

The goal here is to derive the boundary deformation flow equation associated with the RGF. Remarkably, we find that starting the flow from a non-gravitational boundary theory results in the emergence of $4d$ Einstein gravity on the boundary. We refer to this phenomenon as ``\emph{RG induced gravity}''. 

We formulate finite cutoff holography within the saddle-point approximation, where the on-shell actions of the bulk and boundary theories are identified. Our goal is therefore to determine the radial evolution of the bulk on-shell action. To achieve this, we note that \eqref{on-shell-var-action-D} yields 
\begin{equation}\label{delta-xi-action}
    \delta_\xi S\!_r[g]\Big|_{\text{on-shell}} 
    = -\frac{1}{2} \int_{\Sigma(r)} \sqrt{-h}\, T^{ab}\, \delta_\xi h_{ab}\, .
\end{equation}
Taking $\xi$ to be a diffeomorphism along radial direction, $\xi = \partial_{r}$ and using the identity $\delta_{\xi} X = \partial_r X$ for any covariant bulk or boundary quantity $X$, we find~\cite{Parvizi:2025wsg}, (the proof is reviewed in Appendix~\ref{sec:proof})
\begin{equation}\label{TTbar-deformation-flow-NR-1}
    r \frac{\d{}}{\d{r}}   {S}\!_{r}[g] = -  \kappa_5\, L \int_{\Sigma(r)} \sqrt{-h}\,   {\Ottbar}\, . 
\end{equation}
On the other hand, recall that the finite-distance holography proposal within the saddle-point approximation implies \cite{Parvizi:2025wsg}
\begin{equation}\label{Finite-cutoff-Holography-D-NR}
      {S}\!_r[g] \Big|_{\text{on-shell}} \equiv   S_{\text{bdy}} \Big|_{\Sigma(r)} \, ,
\end{equation}
where the right-hand side denotes the boundary action defined on $\Sigma(r)$, which is dual to the bulk theory on $\mathcal{M}(r)$. Thus, we arrive at the desired result:
\begin{equation}\label{TTbar-deformation-flow-NR}
    { r \frac{\d{}}{\d r}   {S}_{\text{bdy}}\Big|_{\Sigma(r)} = -  \kappa_5\, L \,\int_{\Sigma(r)} \sqrt{-h}\,   {\Ottbar}\, . }
\end{equation}

The above is the deformation flow equation for the boundary theory, with $r$ serving as the {RGF} parameter. It is a first-order differential equation that admits a unique solution, once one provides the  boundary (or initial) condition,
\begin{equation}\label{D-bdry-action}
\lim_{r\to \infty}   {S}_{\text{bdy}}\Big|_{\Sigma(r)} = S_{\text{bdy}}\Big|_{\Sigma}\, .
\end{equation}
The right-hand side of \eqref{D-bdry-action} denotes the boundary action evaluated at the asymptotic AdS$_5$ boundary $\Sigma$ defined according to the standard gauge/gravity duality with Dirichlet bc. Eq.~\eqref{TTbar-deformation-flow-NR} captures the central statement that deforming the boundary theory by the $\Ottbar$ operator enables the construction of finite-cutoff holography~\cite{McGough:2016lol, Taylor:2018xcy, Hartman:2018tkw, Parvizi:2025shq, Parvizi:2025wsg, Babaei-Aghbolagh:2024hti, He:2025ppz}.

The first key observation toward our main result is that the deformation flow equation~\eqref{TTbar-deformation-flow-NR} is subject to the Hamiltonian constraint~\eqref{EoM-ss-NR}. Importantly, the $\Ottbar$ operator also appears in the Hamiltonian constraint, and thus \eqref{TTbar-deformation-flow-NR} can be rewritten as, 
\begin{equation}\label{deform-flow-Grav-NR}
    \inbox{ \frac{\d{}}{\d r}   {S}_{\text{bdy}} =  \frac{r}{L\, \kappa_5} \int_{\Sigma(r)} \sqrt{-\gamma} \left[  R[\gamma] + \frac{12}{L^2} \left(\frac{r}{L} \right)^{2} \right]. }
\end{equation}
Note that \eqref{deform-flow-Grav-NR} is expressed in terms of the boundary conformal metric $\gamma_{ab}$, as defined in~\eqref{rescale-1} 

One can integrate \eqref{deform-flow-Grav-NR} using the  Wilsonian EFT formalism, once one specifies the wave-function renormalization equation for the metric $\gamma_{ab}(x^a; r)$, which follows from bulk Einstein's equations. Doing so, we obtain the boundary action at energy scale $\mu$ (see appendix \ref{sec:app-C} for the details of derivation) 
\begin{equation}\label{4d-induced-gravity}
 \hspace*{-2mm} \inbox{\hspace*{-4mm}\begin{split}
     {S}_{\text{bdy}}\Big|_{\Sigma(r)} = & {S}_{\text{bdy}}\Big|_{\Sigma_0}  + \frac{1}{2\kappa_4(\mu)} \int_{\Sigma(r)} \sqrt{-\gamma}   \Bigg[ R[\gamma] -2 \Lambda_4 (\mu) \hspace*{-4mm}\\
         &+{\kappa_4(\mu)\beta(\mu) W_{abcd}[\gamma] W^{abcd}[\gamma]} \Bigg]  +\mathcal{O}(\mu^{-4}), 
    \end{split} \hspace*{-2mm}}
\end{equation}
{where $W_{abcd}[\gamma]$ is the Weyl curvature of metric $\gamma_{ab}$ and }
\begin{equation}\label{4d-grav-coeff}
\inbox{   \begin{split}
       \mu:=\frac{r}{L^2}, \qquad & \kappa_4(\mu) = \frac{2 \kappa_5 }{L^3} \frac{1}{\mu^2}\, , \\ \Lambda_4(\mu) := 6 \mu^2 \left( \frac{\mu_0^4}{\mu^4} -1\right),\quad 
        &\beta(\mu) =  \frac{2 \kappa_5 }{4L^3} \ln \left(\frac{\mu}{\mu_0} \right)\, .
   \end{split}}
\end{equation}
{Here} $\Sigma_0=\Sigma(r_0)$, $\kappa_4 (\mu)$ is the effective induced $4d$ Newton constant, $\Lambda_4(\mu)$ is the effective $4d$ cosmological constant and ${\gamma}_{ab}(x^a; \mu)$ is the renormalized $4d$ gravitational metric. The $\beta$-term denotes the logarithmic corrections to the {Einstein gravity by the seminal conformal gravity \cite{Stelle:1976gc, Adler:1982ri, Mannheim:2011ds, tHooft:2011aa, Maldacena:2011mk}}.  Note that the effective $4d$ gravity couplings \eqref{4d-grav-coeff} have been obtained by assuming the following boundary conditions on the RGF equations: $\kappa_4=0, \Lambda=0, \beta=0$ at $\Sigma_0$ (large $\mu_0\gtrsim\mu$).

The above result is quite nontrivial and requires some discussion and comments.  
Recall that the gauge/gravity correspondence is valid in the large $N$ limit, where the AdS$_5$ gravity describes $4d$ SU$(N)$ gauge theory. In this correspondence the coefficient ${\kappa_5}/{L^3}$ behaves as $1/N^2$ \cite{Aharony:1999ti}.  Parametric smallness of the induced/effective $4d$ Newton constant reassures that our effective gravity description is a valid one, and its $1/N^2$ behavior reaffirms the fact that the $4d$ gravity is induced from the loop effects of the $4d$ gauge theory at low energies. 

While we arrived at the above result using a holographic analysis and through a $5d$ gravity description, the above could be completely discussed without using the $5d$ gravity and only within a $4d$ gauge theory and {RGF} there: We start with a non-gravitating $4d$ gauge theory in the UV, flowing to the IR, among the irrelevant couplings turned on, there is $\Ottbar$, cf. \eqref{TTbar-NR} and \eqref{TTbar-deformation-flow-NR}, which is universal to all $4d$ QFTs and shows up as a $4d$ gravity theory. 
The above is already beyond the usual Wilsonian EFT picture, because the $4d$ metric is now viewed as an independent dynamical field. The $5d$ holographic description prompts the idea that there is no fundamental reason not to allow for the $4d$ metric to be turned on as an effective $4d$ dynamical field. The fact that this is not only a valid viewpoint, but indeed is required, can be understood within the holographic framework, as we discuss in the next section.

\section{Renormalization flow of boundary conditions}

For the above picture of RG-induced gravity to hold, one must relax the Dirichlet bc on the metric field $h_{ab}$ at arbitrary $r$ in the dual holographic description. Imposing $\delta h_{ab} = 0$ on $\Sigma(r)$ effectively turns off $4d$ gravity. As discussed in \cite{Parvizi:2025wsg}, one can impose any desired bc on a given hypersurface $\Sigma(r)$ by adding appropriate boundary terms to the $5d$ gravitational action. However, once a specific boundary term is fixed on $\Sigma(r)$, the bc on any other surface $\Sigma(r')$ is determined by that choice. See \cite{Parvizi:2025wsg, Parvizi:2025shq} for further details.

We explore how the Dirichlet bc imposed at the boundary hypersurface $\Sigma$, specified by $\delta \gamma_{ab}\big|_{\Sigma} = 0$, should evolve as one moves into the bulk onto a hypersurface located at a large but finite radial coordinate $r$. To this end, we perform an asymptotic expansion about the boundary by introducing the dimensionless parameter $\epsilon = L^2 / r^2 \ll 1$, and solve \eqref{EoM-GR-mat-decompose-NR} perturbatively in powers of $\epsilon$. This yields (see Appendix~\ref{sec:asymp-exp} for details)
\begin{equation}\label{brdy-condn-RG-flow}
    \inbox{\delta h_{ab}(\epsilon) = -\frac{\kappa_5}{5 L^3} \, \delta \left( L^4 T_{ab}(\epsilon) \right) + \mathcal{O}(\epsilon^2) \, , }
\end{equation}
where $L^4 T_{ab}$ is the dimensionless BY-EMT. Thus, the Dirichlet bc originally imposed at $\Sigma$ effectively evolves into a mixed Dirichlet–Neumann bc at small but finite $\epsilon$.
The appearance of ${\kappa_{5}}/{L^3}\sim 1/N^2$ factor indicates that the evolution may be treated perturbatively in the large $N$ limit.

The above construction can be extended to move finitely away from $\Sigma$. Let us denote the boundary data on $\Sigma$ and $\Sigma(r)$ by $\{h_{ab}^{\infty},   {  {T}}_{\infty}^{ab}\}$ and $\{h_{ab}(r), {  {T}}^{ab}(r)\}$, respectively. One can in principle solve  \eqref{EoM-GR-mat-decompose-NR} to obtain 
\begin{equation}
    h_{ab}^{\infty}=h_{ab}^{\infty}[h_{ab}(r),  {T}^{ab}(r)]\, , \quad   {T}_{\infty}^{ab}=  {T}_{\infty}^{ab}[h_{ab}(r),     {T}^{ab}(r)]\, .
\end{equation}
While the above explicit solution may not be analytically available, the solution should exist \cite{Aharony:1999ti, Skenderis:2002wp}. Thus, the bc on $\Sigma(r)$ is \textit{induced} from the asymptotic Dirichlet bc:
\begin{equation}\label{mixed-bc-Sigma-r}
   \inbox{\hspace*{-2mm} \text{bc on $\Sigma(r)$:} \quad   \delta h_{ab}^{\infty}[h_{ab}(r),    {T}^{ab}(r)]  = 0\, , \hspace*{-2mm}}
\end{equation}
yielding a mixed Neumann and Dirichlet bc's at generic $r$.

\section{Outlook}

We have uncovered a remarkable outcome of the holographic  correspondence: 
\begin{center}
\textit{Gravity is not a fundamental force; it is an effective IR description of deformations by certain irrelevant operators. These deformations are universal to all QFTs, yielding to the universal feature of gravity.} \end{center}
While we used $5d$  holographic setting for our computations and analysis, the statements and the results can be stated only within the $4d$ theory and its {RGF}, without recrossing to $5d$. 

Our result indicates that the decoupling of scales and Wilsonian EFT description should break down at some particular energy scale. This feature is already expected from the AdS/CFT duality: Existence of a gravity description in the bulk and the fact that one can move both ways, up or down, in the radial direction means that we should be able to ``integrate in'' degrees of freedom as well as the usual integrate-out and coarse-graining \cite{Verlinde:2010hp, Sheikh-Jabbari:2013AdSCFT}. This is related to another known fact in the AdS/CFT that locality of the boundary theory does not imply locality of the bulk theory, and vice-versa \cite{Heemskerk:2009pn, Almheiri:2014lwa}.

The RG-induced gravity arises from two features: (1) \textit{{RGF} in boundary conditions, \eqref{brdy-condn-RG-flow} and \eqref{mixed-bc-Sigma-r}.} As we fix the Dirichlet bc on the AdS boundary (and hence start with a non-gravitating theory in the UV), the bc for the metric at any radius inside AdS is a mixture of Neumann and Dirichlet, and hence the boundary metric away from the UV admits non-trivial fluctuations. (2) \textit{Induced Einstein gravity.} Variation of the on-shell bulk action involves a $\Ottbar$ term, \textit{cf.} \eqref{TTbar-deformation-flow-NR}. Upon the gauge/gravity correspondence and recalling the $5d$ ``Hamiltonian constraint'' \eqref{EoM-ss-NR}, this may be replaced with an effective $4d$ Einstein gravity plus curvature squared corrections \eqref{4d-induced-gravity}.

\usetikzlibrary{decorations.markings} 

\usetikzlibrary{decorations.markings} 

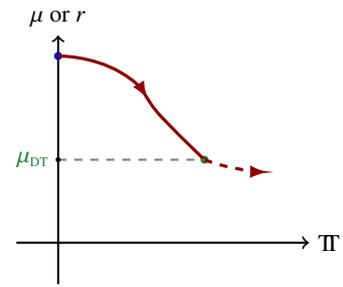
\begin{figure}[t]
    \centering
    \begin{tikzpicture}[scale=.55, thick, decoration={markings, mark=at position 0.55 with {\arrow{latex}}}]
    \draw[->] (-1,0) -- (6,0) node[right] {$\Ottbar$};
    \draw[->] (0,-1) -- (0,5) node[above] {$\mu$ or $r$};

    \filldraw[blue!80!black] (0,4.5) circle (2pt);

    \filldraw[darkgreen] (3.5,2) circle (2pt);

    \draw[very thick, darkred!80!black, postaction={decorate}] 
        plot [smooth, tension=1] coordinates {(0,4.5) (1.5,4.1) (2.5,3.0) (3.5,2)};

    \draw[very thick, darkred!80!black, dashed, ->, >=latex] 
        plot [smooth, tension=1.1] coordinates {(3.5,2) (4.2,1.8) (5,1.7)};

    \draw[dashed, gray] (0,2) -- (3.5,2); 
    \filldraw[black] (0,2) circle (1pt) node[left] {\textcolor{darkgreen}{$\mu_{\text{\tiny{DT}}}$}};

    \end{tikzpicture}
    \caption{ \justifying{\footnotesize {RGF} in the boundary theory, starting from a UV CFT on a flat background. Along the flow, and for $\mu>\mu_{\text{\tiny{DT}}}$ with the transmutation scale $\mu_{\text{\tiny{DT}}}$, the theory admits two equivalent interpretations: a QFT deformed by the $\Ottbar$  operator, or a gravitating boundary theory. For $\mu<\mu_{\text{\tiny{DT}}}$ the bc effectively becomes Neumann and only the gravitational interpretation remains valid. } }
    \label{fig:RGflow}
\end{figure}

We differ from the  T$\bar{\text{T}}$ deformation extensively discussed in the recent literature \cite{McGough:2016lol, Taylor:2018xcy, Hartman:2018tkw, Guica:2019nzm}, where the bc at the finite cutoff surface $\Sigma(r)$ is always set to be Dirichlet. With this choice, the boundary theory on $\Sigma(r)$ is non-gravitating and the effects of the {RGF} from $\Sigma$ to $\Sigma(r)$ are only encoded in the deformation of the theory by the T$\bar{\text{T}}$ operator. 

Eq.~\eqref{brdy-condn-RG-flow} may be viewed as the beta-function equation for evolution in bc's.  In principle, one can integrate out this beta-function. The Neumann admixture in the mixed bc is expected to increase as we decrease $r$ and one expects in the deep IR to effectively find a Neumann bc $\delta (\sqrt{-h} T^{ab})=0$, while $\delta h_{ab}$ remains unconstrained, as expected in a standard $4d$ gravity theory. It is conceivable that we find a phenomenon similar to the dimensional transmutation and spontaneous generation of scale that we are familiar with in usual gauge theories \cite{Coleman:1973jx, Weinberg:1996kr}, we find a scale $\mu_{\text{\tiny{DT}}}$ where the bc is predominantly Neumann (see Fig.\ref{fig:RGflow}). Above $\mu_{\text{\tiny{DT}}}$, we have the freedom to interpret the deformations induced through the RGF in gauge theory as a T$\bar{\text{T}}$ deformation with bc's viewed as a deformation of Dirichlet, or alternatively as an induced gravity with bc's viewed as a deformed Neumann. At or below $\mu_{\text{\tiny{DT}}}$, however, the T$\bar{\text{T}}$ deformation interpretation is not available, and one should use the effective induced $4d$ gravity description. 

In the analysis of the main text, we considered an unrenormalized holographic {RGF} setting. However, it is known that the on-shell gravity/boundary action is divergent. In the holographic renormalization program \cite{deBoer:1999tgo, deHaro:2000vlm}, appropriate counterterms have been introduced to remove the divergences \cite{Balasubramanian:1999re, Emparan:1999pm, Skenderis:2002wp}. In the appendix \ref{sec:RIG}, we have extended our analysis to the renormalized case. One finds a ``holographically renormalized induced $4d$ gravity'' which, compared to the unrenormalized case, has the following remarkable features: (1) Effective (renormalized) $4d$ gravity coupling is now becoming a constant, with no scale $\mu$ dependence. (2) The effective (renormalized) cosmological constant is zero, prompting some ideas to resolve the cosmological constant problem.

We close with a comment that besides Einstein gravity, \eqref{4d-induced-gravity} has a logarithmic correction by the {Weyl-squared (conformal gravity) term \cite{Stelle:1976gc, Adler:1982ri, Mannheim:2011ds, tHooft:2011aa, Maldacena:2011mk}.}  
There are other higher-order corrections to \eqref{4d-induced-gravity},  involving third-order or higher powers or higher derivatives of curvature. There are also corrections involving products of second or higher powers of curvature and energy-momentum tensor of the matter fields coupled to the effective $4d$ gravity. Our holographic framework allows for a systematic derivation of these corrections.  See appendices \ref{sec:app-C} and \ref{sec:RIG} for a sketch of the derivation.

\section*{Acknowledgement}
We would like to thank Dionysios Anninos, Daniel Grumiller, Juan Maldacena, and Steve Shenker for comments on the manuscript. MMShJ acknowledges BIMSA and Hossein Yavartanoo for the hospitality, where he gave a minicourse on the Freelance Holography Program, and Sergio Cecotti for a comment in the minicourse that sparked this work. The work of HA is supported by the Beijing Natural Science Foundation under Grant No. IS23018. The work of VT is supported by the Iran National Science Foundation (INSF) under project No. 4040771.

\appendix
\section*{\large{Supplemental Material}}
In the appendices \ref{sec:details}, \ref{sec:asymp-exp}, and \ref{sec:app-C}, we provide details of derivations. In the main text we discussed the gravity induced on the boundary within an unregularized holographic {RGF} analysis. However, one could have used the regularized holographic {RGF} \cite{Skenderis:2002wp,Balasubramanian:1999re, Emparan:1999pm}. 

In appendix \ref{sec:RIG} we discuss the regularized case and derive the ``holographically renormalized RG-induced gravity theory. We show that in this case, (1) the effective induced $4d$ gravity does not involve a cosmological constant term and, (2) the effective induced $4d$ coupling is a constant (with no scale dependence). The former may provide a resolution to the cosmological constant problem: the zero-point energies in a renormalized holographically induced gravity theory do not gravitate. The latter  shows that how the non-renormalizability of the usual $4d$ Einstein gravity (and that the effective dimensionless gravitational coupling scales like $\mu^2$, as is also seen in \eqref{4d-induced-gravity} and \eqref{4d-grav-coeff}), is removed/addressed in the renormalized case. 
\section{Details of some derivations }\label{sec:details}
\textbf{More on Eq.~\eqref{EoM-GR-mat-decompose-NR}.} We first note that \eqref{Tab-def} yields
\begin{equation}\label{T[h]}
    \sqrt{-h} \, T^{ab} = -\frac{1}{2\kappa_5 L} \, \frac{1}{\sqrt{-h}}\, r\partial_r \left( -h\, h^{ab} \right)\,,
\end{equation}
i.e., $T^{ab}$ can be regarded as a functional of the induced metric $h_{ab}$,  $T^{ab} = T^{ab}[h_{cd}]$. 
From \eqref{EoM-ab-NR} and \eqref{Tab-def} we learn that 
\begin{equation}\begin{split}
r\partial_r \left(\sqrt{-h}  {  {T}}^{ab}\right) &=
      \sqrt{-h}  \Bigg[\kappa_5 L\left(-2   {  {T}}^{a}_{\;c}   {  {T}}^{bc} 
       + \frac{2}{3}   {  {T}}   {  {T}}^{ab} 
        \right) \\
        & \hspace{1 cm}+\frac{L}{\kappa_5}  \left(R^{ab}-  R h^{ab} -\frac{12}{ L^2}h^{ab}\right)\Bigg]\, , \label{EoM-ab-UP}\\
   r \partial_r \left(\sqrt{-h}\,   {  {T}} \right) &= -\sqrt{-h}\left(\frac{48}{\kappa_5 L} + \frac{3L}{\kappa_5} R \right),\\
    r \partial_r \ln\sqrt{-h} &=-\frac{L\kappa_5}{3}   {T}\, .
\end{split}
\end{equation}

Eq.~\eqref{EoM-ss-NR} implies, 
\begin{equation}
\begin{split}
&\frac{1}{\sqrt{-h}}\frac{\delta}{\delta h_{ab}} \Big[\sqrt{-h} \Big(R+\frac{12}{L^2}+\kappa_5^2  {\Ottbar} \Big) \Big] =0 \quad \Longrightarrow \\
&\frac{\delta   {\Ottbar}}{\delta   {  {T}}^{cd}}\frac{\delta  {T}^{cd}}{\delta h_{ab}} +  2   {  {T}}^{ac}   {  {T}}^{b}{}_c 
       - \frac{2}{3}   {  {T}}  {  {T}}^{ab} =\frac{1}{\kappa_5^2} R^{ab}\, ,
\end{split}  
\end{equation}
and from \eqref{TTbar-NR} we have 
\begin{equation}\label{TTbar-variation}
    \frac{\delta   {\Ottbar}}{\delta   {  {T}}^{cd}}= 2 (  {  {T}}_{cd} - \frac{1}{3}   {  {T}} h_{cd})= \frac{1}{\kappa_5 L} r\partial_r h_{cd}\, .
\end{equation}
Therefore,  \eqref{EoM-sa-NR} and \eqref{EoM-ab-NR} may be written as,
\begin{subequations}\label{r-der-Tab}
\begin{align}
    r\partial_r \left(\sqrt{-h}\, T^{ab}\right) - \frac{\delta (\sqrt{-h}\, {T}^{cd})}{\delta h_{ab}}\, r\partial_r h_{cd}&=0\,, \\
    \nabla_a \left(\sqrt{-h}\, T^{ab}\right) = \frac{1}{2\kappa_5 L} \, \frac{1}{\sqrt{-h}}\,[r\partial_r,\nabla_a] \left( -h\, h^{ab} \right) &= 0\,.\label{div-T-h}
\end{align}
\end{subequations}
The above should be solved together with  \eqref{EoM-ss-NR}. In deriving the above, we used the fact that 
\begin{equation}
       {\Ottbar}=   {  {T}}^{cd}(  {  {T}}_{cd}-\frac13   {  {T}} h_{cd})= \frac{1}{2\kappa_5 L}  {  {T}}^{cd} r\partial_r h_{cd}\, .
\end{equation}
Eq.\eqref{r-der-Tab} is a  consistency check for the fact that  $ {T}^{ab}$, on-shell, can be treated as functional of  $h_{ab}$, ${T}^{cd}={T}^{cd}[h_{ab}]$. This can also be seen explicitly in \eqref{T[h]}.

As the final comment, we note  that $ {\Ottbar}$ can be written as
\begin{equation}
(2L\kappa_5)^2   {\Ottbar}= (r\partial_r h_{ab})(r\partial_r h_{cd}) {\cal G}^{abcd}\, ,    
\end{equation}
where ${\cal G}^{abcd}$ is the WdW metric \cite{PhysRev.160.1113},
\begin{equation}
    {\cal G}^{abcd}=h^{ac}h^{bd}+h^{ad}h^{bc}-2h^{ab}h^{cd}\, .
\end{equation}

\textbf{Proof of \eqref{TTbar-deformation-flow-NR-1}.}\label{sec:proof}  
We start with \eqref{delta-xi-action}, take $\xi = \partial_r$ and use the identity $\delta_{\partial_r} X = \partial_r X$ for any covariant bulk or boundary quantity. Then,
\begin{equation}  
   \begin{split}
     \hspace{-.12 cm}  \frac{\d{}}{\d{}r}   S\!_r[g] \Big|_{\text{on-shell}} & = -\frac{1}{2} \int_{\Sigma(r)} \sqrt{-h}\,  {  {T}}^{ab}\, \partial_{r} h_{ab} \\
       & = - \frac{L}{r}\int_{\Sigma(r)} \sqrt{-h}\,   {  {T}}^{ab}\, K_{ab} \\
       & = - \frac{ \kappa_5 L}{r}\int_{\Sigma(r)} \sqrt{-h}  {  {T}}^{ab} \left(  {  {T}}_{ab} - \frac{  {  {T}}}{3} h_{ab} \right) \\
       & = - \frac{ \kappa_5 L}{r} \int_{\Sigma(r)} \sqrt{-h}\,   {\Ottbar}\, .
   \end{split}
\end{equation}
In the second line, we employed the definition of the extrinsic curvature given in \eqref{def-K}, and in the third line, we expressed the extrinsic curvature in terms of the BY-EMT using \eqref{BY-EMT}. Finally, in the last line, we used the definition of $\Ottbar$ in \eqref{TTbar-NR}.
\section{Asymptotic expansion}\label{sec:asymp-exp}
The on-shell asymptotic (large-$r$) expression of $\gamma_{ab}$ is given by \cite{deHaro:2000vlm}
\begin{equation}\label{gamma-epsilon}
     \begin{split}
         \gamma_{ab}(\epsilon) &= \gamma^\infty_{ab} - \epsilon\, S_{ab}[\gamma^\infty]+ \epsilon^2 \ln \epsilon^2\, \tilde S_{ab}[\gamma^\infty] \\ 
         & - \frac{\epsilon^2}{2} \Big( \frac{\kappa_{5}}{L^3} \hat{T}^{\infty}_{ab} + \hat{S}_{ab} [\gamma^\infty] \Big) + \mathcal{O}(\epsilon^3) \, ,
     \end{split}
\end{equation}
where $\epsilon := \frac{L^2}{r^2}$ and $S_{ab}[\gamma^\infty]$ denotes the dimensionless Schouten tensor associated with 
$\gamma_{ab}^{\infty}$
\begin{equation}\label{Schouten-def}
    S_{ab}[\gamma^\infty] := \frac{L
    ^2}{2} \Big( R^{\infty}_{ab} - \frac{1}{6} R^{\infty} \gamma_{ab}^{\infty} \Big)\, .
\end{equation}
The tensors $\tilde S_{ab}[\gamma^\infty]$ and $\hat{S}_{ab}[\gamma^\infty]$ are defined, respectively, as
\begin{equation}
    \begin{split}
         & \tilde S_{ab}[\gamma^\infty]  = \frac{L^2}{16} \Big[ 4 L^{-2} S_{a}^{c}S_{bc} - 2 D_{c} D_{(a} S_{b)}^{c} + D_{a} D_{b} S \\
        & \hspace{.5 cm} + \Box S_{ab} + \Big( \frac{3}{2} D_{c} D_{d} S^{cd} - \frac{3}{2} \Box S - L^{-2} S^{cd} S_{cd} \Big)\gamma_{ab}^{\infty}  \Big]\, , \\
        & \hat{S}_{ab}[\gamma^\infty] = - S_{a}^{c} S^{cb} +\frac{1}{2} S S_{ab} + \frac{1}{4}(S_{cd}S^{cd} -S^2) \gamma^{\infty}_{ab}\, ,
    \end{split}
\end{equation}
where $D_a$ is the covariant derivative compatible with the asymptotic boundary metric $\gamma_{ab}^{\infty}$, and $\Box := D_a D^a$. Note that 
$\gamma_{ab}^\infty \tilde S^{ab}=0$. The above yields, 
\begin{equation}\label{T-epsilon}
    \begin{split}
    \hspace{-0.3 cm}  \frac{\kappa_5}{L^3} (L^4{T}_{ab}(\epsilon)) &=  -3 \epsilon^{-1} \gamma_{ab}^{\infty} +(4 S_{ab} - S \gamma_{ab}) - 10\epsilon \ln \epsilon\, \tilde{S}_{ab} \\
        &+\epsilon \Big[ \frac{5 \kappa_5}{2 L^3} \hat{{T}}^{\infty}_{ab} - 2 \tilde{S}_{ab} - \frac{5}{2} S_{a}^{c}S_{cb} + \frac{9}{4} S S_{ab} \\
        &+\frac{1}{8} \Big( S_{cd} S^{cd} -5 S^2 \Big) \gamma^{\infty}_{ab} \Big] + \mathcal{O}(\epsilon^2)\, .
    \end{split}
\end{equation}
The asymptotic solution space is parametrized by two codimension-one tensors, $\{\gamma_{ab}^{\infty},  \hat{{T}}^{\infty}_{ab}\}$, which are subject to the following constraints
\begin{equation}
    D_{a} \hat{{T}}_{\infty}^{ab} = 0\, , \qquad  \hat{{T}}^{\infty} = -\frac{L^3}{2 \kappa_5} (S^{ab} S_{ab} - S^{2})\, .
\end{equation}

As noted in the main text, we impose the Dirichlet boundary condition at infinity, $\delta \gamma_{ab}^{\infty} = 0$. Varying  \eqref{gamma-epsilon} and \eqref{T-epsilon}, we obtain the corresponding induced bc at finite $\epsilon$
\begin{equation}\label{delta-gamma-epsilon-1}
     \begin{split}
        & \delta \gamma_{ab}(\epsilon) =  - \frac{\epsilon^2}{2} \frac{\kappa_{5}}{L^3} \delta \hat{{T}}^{\infty}_{ab} +\mathcal{O}(\epsilon^3)\, , \\
        &\delta (L^4{T}_{ab}(\epsilon)) = \frac{5\epsilon}{2} \delta \hat{{T}}^{\infty}_{ab} + \mathcal{O}(\epsilon^2)\, .
     \end{split}
\end{equation}
Combining these equations leads to
\begin{equation}
    \delta \gamma_{ab}(\epsilon) =  - \frac{\epsilon}{5} \frac{\kappa_{5}}{L^3} \delta \left( L^4 {T}_{ab}(\epsilon) \right) +\mathcal{O}(\epsilon^3)\, ,
\end{equation}
reproducing the boundary condition flow \eqref{brdy-condn-RG-flow}.
\section{Holographically induced gravity action}\label{sec:app-C}
Now we aim to solve the  RGF equation \eqref{deform-flow-Grav-NR},
\begin{equation}\label{deform-flow-Grav-NR-1}
    \frac{\d{}}{\d r}   {S}_{\text{bdy}} =  \frac{r}{L\, \kappa_5} \int_{\Sigma(r)} \sqrt{-\gamma} \left[  R[\gamma] + \frac{12}{L^2} \left(\frac{r}{L} \right)^{2} \right],
\end{equation}
perturbatively in the large-$r$. This equation may be written as
\begin{equation}\label{4d-action-flow}
   \hspace{-2.5 mm}   S_{\text{bdy}}\Big|_{\Sigma(r+\d r)} = S_{\text{bdy}}\Big|_{\Sigma(r)} + \frac{r\, \d r}{L\, \kappa_5 } \int_{\Sigma(r)} \sqrt{-\gamma} \left[ {R}[\gamma] + \frac{12}{L^2} \left(\frac{r}{L} \right)^{2} \right].
\end{equation}

To solve the above differential equation, consider the following solution ansatz: 
\begin{equation}\label{soln-ansatz}
\begin{split}
      {S}_{\text{bdy}}\Big|_{\Sigma(r)} = & {S}_{\text{bdy}}\Big|_{\Sigma_0}  + \frac{1}{2\kappa_4(r)} \int_{\Sigma(r)} \sqrt{-\gamma}   \Bigg[ R[\gamma] -2 \Lambda_4 (r) \hspace*{-4mm}\\
       &+\kappa_4(r)\beta(r) \, W_{abcd}[\gamma]W^{abcd}[\gamma]  \Bigg]  +\mathcal{O}(r^{-4})\, , 
\end{split}\end{equation}
and solve for the unknown coefficient $\kappa_4(r), \Lambda_4(r), \beta(r)$. {Note that in writing the above, we used the identity, 
\begin{equation}\label{W2-GB-critical}\begin{split}
    W_{abcd}W^{abcd} &= 2 \Big(R_{ab} R^{ab} - \frac{1}{3} R^2 \Big) + \text{GB}\, , \\
    \text{GB} &=R_{abcd}R^{abcd}-4 R_{ab}R^{ab}+R^2\, , 
\end{split}\end{equation}
where GB is the Gauss-Bonnet term which is a topological (total derivative) term in $4d$ and thus do not contribute to equations of motion.} 
In the above $ {S}_{\text{bdy}}\Big|_{\Sigma_0}$is an integration constant, for a boundary at a large $r=r_0$, and we solve the equation for $r<r_0$. To solve this equation, we need to have the information of $\gamma_{ab}(r)$. Recalling the general form of $5d$ Einstein equations, we have:
\begin{equation}\label{Ricci-flow-eq}
    \partial_r \gamma_{ab}=\frac{2}{L}\left(\frac{L}{r}\right)^3 {S_{ab}[\gamma]}+ {\cal O}(r^{-5})\, , 
\end{equation}
where ${S_{ab}[\gamma]}$ is the dimensionless Schouten tensor. The above is the leading part of the exact equation \eqref{gamma-ab-wave-funcn-renorm}. In the $4d$ theory, the above is to be viewed as the wave-function renormalization for the metric field $\gamma_{ab}(x^a; r)$. In a different viewpoint, \eqref{Ricci-flow-eq} may be viewed as a Ricci (Schouten) flow equation. 

Inserting \eqref{soln-ansatz} into \eqref{deform-flow-Grav-NR-1}, we obtain
\begin{equation}\begin{split}\label{RG-eq-conpuling}
 \frac{\d{}}{\d{}r} \left(\frac{\Lambda_4}{\kappa_4}\right)&=-\frac{12}{L^2\kappa_5} \left(\frac{r}{L}\right)^3\, , \\
    \frac{\d{}}{\d{}r}\left(\frac{1}{\kappa_4}\right)&=\frac{2r}{L\kappa_5}+ \frac{L}{3}\left(\frac{\Lambda_4}{\kappa_4}\right)\left(\frac{L}{r}\right)^3\, ,\\
     \frac{\d{}\beta}{\d{}r}&= \frac{L}{2\kappa_4}\left(\frac{L}{r}\right)^3\, .
\end{split}
\end{equation}
One can simply integrate these equations to obtain 
\begin{equation}\begin{split}
\frac{\Lambda_4(r)}{\kappa_4(r)}&=-\frac{3}{L\kappa_5} \left(\frac{r}{L}\right)^4 +C_0\, , \\
    {\kappa_4(r)}&=\frac{2\kappa_5}{L}{\left(\frac{L}{r}\right)^2}\, ,\\
     \beta(r)&= \frac{L^3}{4\kappa_5}\ln\left(\frac{r}{r_0}\right)\, ,\\
\end{split}
\end{equation}
where $r_0$ and $C_0$ are integration constants. Note that the above are the leading-order solutions in an $r/L$ expansion. {It is worth emphasizing that, in  \eqref{RG-eq-conpuling}, we have discarded the subleading cubic curvature terms as well as quadratic curvature terms with two derivatives.}

\section{Holographically Renormalized Induced Gravity}\label{sec:RIG}
In the main text, we considered holography with unrenormalized theories on both sides. In this section, we work with renormalized theories and rederive the holographically renormalized RG-induced gravity theory.

\subsection{Renormalized action principle} The $5d$ bulk Einstein–Hilbert action compatible with Dirichlet boundary conditions on $\Sigma(r)$ is given in \eqref{action-AdS-NR}. The on-shell action diverges in the limit $r \to \infty$. To obtain a renormalized on-shell action,\footnote{It is better to call this regularized (rather than renormalized) on-shell action and similarly in most places, holographic regularization (rather than holographic renormalization). We follow the commonly used terminology, and use ``renormalized'' instead of the more precise term ``regularized''.} we begin with \cite{Balasubramanian:1999re, Emparan:1999pm} 
\begin{equation}\label{action-AdS}
     {S}^{\text{\tiny{Ren}}}_r[g] = {S}\!_r[g] + \frac{L}{4\kappa_5} \int_{\Sigma(r)} \sqrt{-h} \left(  \frac{12}{L^2} + R[h]\right)\, .
\end{equation} 
The last term in the action above is a boundary term consisting of counterterms \cite{Balasubramanian:1999re, Emparan:1999pm}, which do not affect the Dirichlet boundary condition, as they are only functionals of $h_{ab}$, and are introduced to render the on-shell action finite.

The on-shell variation of the renormalized Einstein–Hilbert action \eqref{action-AdS} gives
\begin{equation}
    \delta {S}^{\text{\tiny{Ren}}}_r[g]  \Big|_{\text{on-shell}} = -\frac{1}{2} \int_{\Sigma(r)} \sqrt{-h}\,   \mathcal{T}^{ab}\, \delta h_{ab}\, , 
\end{equation}
where \( \mathcal{T}^{ab} \) is the renormalized Brown–York energy-momentum tensor (rBY-EMT), also known as the renormalized holographic energy-momentum tensor \cite{Balasubramanian:1999re}. It consists of standard BY-EMT \eqref{BY-EMT} and counterterm energy-momentum tensor \cite{Balasubramanian:1999re, Emparan:1999pm}
\begin{equation}\label{r-BY-EMT}
\begin{split}
      \mathcal{T}^{ab} &=   {  {T}}^{ab}+  {T}_{\text{\tiny{ct}}}^{ab}\, ,\\    
      {T}_{\text{\tiny{ct}}}^{ab}  & 
      =  -\frac{1}{L\kappa_5} \big( {S}^{ab}-S h^{ab} -3 h^{ab}  \big)\, ,   
\end{split}
\end{equation}  
where $S_{ab}=S_{ab}[h]$ is the dimensionless Schouten tensor \eqref{Schouten-def} of the metric $h_{ab}$. The Einstein field equations \eqref{EoM-GR-mat-decompose-NR} in terms of the rBY-EMT take the form 
\begin{subequations}\label{EoM-GR-mat-decompose}
    \begin{align}
        &  \kappa_{5}^2\, \cttbar + 2\, \frac{ \kappa_{5}}{L}\left({\mathcal{T}} + S^{ab} \mathcal{T}_{ab} \right) + L^{-2} (S^{ab}S_{ab} - S^2) = 0\, ,\label{EoM-ss}\\
        & {\nabla}_b  \mathcal{T}^{b}_{a}=0\, , \label{EoM-sa}\\
        & r \partial_{r}    \mathcal{T}_{ab}  = -2   \mathcal{T}_{ab} -    \mathcal{T} h_{ab} + L \kappa_5 \Big( 2    \mathcal{T}_{ac}    \mathcal{T}^{c}_{b} -\frac{1}{3}   \mathcal{T}    \mathcal{T}_{ab} \Big) \nonumber \\
        & \hspace{{1. cm}}+ \frac{L}{2\kappa_5}\Big[ 2 W_{acbd} S^{cd} -L^{-2}(4 S_{ac} S_{b}^{c} - S^{2} h_{ab}) + \Box S_{ab}  \nonumber \\
        & \hspace{{1. cm}} - \nabla_{a}\nabla_{b}S\Big] + \Big[L^2 W_{acbd} \mathcal{T}^{cd} - S \mathcal{T}_{ab} + \frac{2}{3} \mathcal{T} S_{ab} \nonumber\\
        & \hspace{{1. cm}} - \frac{1}{6} \mathcal{T} S  +\frac{L^2}{6} (\nabla_{a} \nabla_{b}\mathcal{T} +3 \Box \mathcal{T}_{ab} - \Box \mathcal{T} h_{ab})\Big]\, ,\label{EoM-ab}
    \end{align}
\end{subequations}  
where $\Box := \nabla_{a} \nabla^{a}$, $W_{acbd}$ is the Weyl tensor and all curvature tensors are constructed from the boundary metric $h_{ab}$. The quantity $\cttbar$ denotes the renormalized $\Ottbar$ operator
\begin{equation}
    \cttbar =   \mathcal{T}^{ab}   \mathcal{T}_{ab} - \frac{1}{3}  \mathcal{T}^2\, .
\end{equation}
The radial evolution of the trace of the rBY-EMT is given by
\begin{equation}\label{EoM-tr}
    \begin{split}
        r \partial_{r}   \mathcal{T} &= -2 \mathcal{T} + L \kappa_{5} \Big(3 \cttbar + \frac{\mathcal{T}^2}{3} \Big) +  (4 S^{ab} \mathcal{T}_{ab} - S \mathcal{T}) \\
        & + \frac{1}{L\kappa_5} (S_{ab} S^{ab} - S^2) \, .
    \end{split}
\end{equation}
\subsection{RG equation for the  boundary metric \texorpdfstring{$\gamma_{ab}(x^c;r)$}{gamma}}
{From the definition of rBY-EMT \eqref{r-BY-EMT}} and \eqref{Tab-def}, we learn that, 
\begin{equation}\label{gamma-ab-wave-funcn-renorm}
  \hspace{-.2 cm}  r \partial_{r} \gamma_{ab}= \frac{2L^2}{r^2} S_{ab}[\gamma] +\frac{2 L^5 \kappa_{5}}{r^4} \left(\hat{\mathcal{T}}_{ab} - \frac{\hat{\mathcal{T}}}{3}\gamma_{ab} \right)\, .
\end{equation}
where, recalling \eqref{rescale-1}, we have introduced,
\begin{equation} \label{rescale-2}
    \hat{\mathcal{T}}_{ab} := \left(\frac{r}{L} \right)^{2} \mathcal{T}_{ab}\, . 
\end{equation}   
In the large-$r$ limit, the second term becomes subleading, and the dominant contribution to the {RGF} of
\begin{equation}\label{gamma-ab-RG-leading}
   r \partial_{r} \gamma_{ab}=\frac{2L^2}{r^2} S_{ab}[\gamma]+ {\cal O}\left(\left(\frac{L}{r}\right)^4\right)\,  .
\end{equation}
This equation closely resembles the celebrated Ricci flow equation, suggesting a geometric interpretation of the boundary {RGF} in terms of a curvature-driven evolution. This is  \eqref{Ricci-flow-eq}, which we used in Appendix \ref{sec:app-C}.

\subsection{Renormalized deformation flow equation} 

We now turn to the derivation of the renormalized deformation flow equation. To this end, we follow the same procedure as in the unrenormalized case
\begin{equation}
   \begin{split}
      \hspace{-3mm} \frac{\mathrm{d}{}}{\mathrm{d}{}r} {S}^{\text{\tiny{Ren}}}_r[g]  \Big|_{\text{on-shell}} & = -\frac{1}{2} \int_{\Sigma(r)} \sqrt{-h}\,{  \mathcal{T}}^{ab}\, \partial_{r} h_{ab} \\
       & = - \frac{ \kappa_5 L}{r} \int_{\Sigma(r)} \sqrt{-h} \left(  \mathcal{T}^{ab}  {T}_{ab} - \frac{1}{3}   \mathcal{T}  {T}\right)\\
       &=- \frac{1}{r} \int_{\Sigma(r)} \sqrt{-h} \left(\mathcal{T}+\mathcal{T}^{ab}S^{ab}+ \kappa_5 L\cttbar \right),\nonumber
   \end{split}
\end{equation}
where in the last line we used \eqref{r-BY-EMT}. 
We now employ the renormalized version of the finite-distance holography proposal in the saddle-point approximation \cite{Parvizi:2025wsg}
\begin{equation}\label{Finite-cutoff-Holography-D}
   {S}^{\text{\tiny{Ren}}}_r[g] \Big|_{\text{on-shell}} = {S}^{\text{\tiny{Ren}}}_{\text{bdy}}\Big|_{\Sigma(r)}\, ,
\end{equation}
where the right-hand side denotes the renormalized boundary action defined on $\Sigma(r)$, which is holographically dual to the renormalized bulk theory on $\mathcal{M}(r)$ \eqref{action-AdS} and is subject to the boundary behavior
\begin{equation}
    \lim_{r\to \infty} {S}^{\text{\tiny{Ren}}}_{\text{bdy}}\Big|_{\Sigma(r)} =  {S}^{\text{\tiny{Ren}}}_{\text{bdy}}\Big|_{\Sigma}\, , 
\end{equation}
where the right-hand side represents the boundary action evaluated at the asymptotic AdS boundary $\Sigma$, defined in accord with the standard gauge/gravity duality.

The above deformation flow equation is constrained by the Hamiltonian constraint \eqref{EoM-ss}. Upon the Hamiltonian constraint, we can recast the flow equation in the following form
\begin{equation}
    \begin{split}
        r \frac{\mathrm{d}}{\mathrm{d}r} {S}^{\text{\tiny{Ren}}}_{\text{bdy}}\Big|_{\Sigma(r)} &=  \int_{\Sigma(r)} \sqrt{-h}\Big[  \mathcal{T} +  S^{ab} \mathcal{T}_{ab} + \frac{1}{L\kappa_5} (S^{ab}S_{ab}-S^2)\Big]\\
        &=  \int_{\Sigma(r)}  \sqrt{-\gamma}\Big[  \hat{\mathcal{T}} + \frac{L^2}{r^2} S^{ab}[\gamma] \hat{\mathcal{T}}_{ab}\\
        &+ \frac{1}{L\kappa_5} \big(S^{ab}[\gamma]S_{ab}[\gamma]-S[\gamma]^2 \big)\Big].
    \end{split}
\end{equation}
where the second equality is expressed in terms of the boundary metric $\gamma_{ab}$ and rescaled rBY-EMT. 

We now solve the above action RGF at leading order in the large-$r$. To proceed, we first express the Hamiltonian constraint in terms of the rescaled variables
\begin{equation}
     \begin{split}
         & \kappa_{5}\, \frac{L^4}{r^4} \hat{\cttbar} + 2\frac{ \hat{\mathcal{T}}}{L} + 2 \frac{L}{r^2} S^{ab}[\gamma] \hat{\mathcal{T}}_{ab} \\
         & + \frac{1}{L^2\kappa_5} \big(S^{ab}[\gamma]S_{ab}[\gamma] - S[\gamma]^2 \big) = 0\, .
     \end{split}
\end{equation}
From this equation, we deduce that at the asymptotic boundary, $\hat{\mathcal{T}}$ satisfies the following relation
\begin{equation}
    \hat{\mathcal{T}} = - \frac{1}{2 L\kappa_5} \big(S^{ab}[\gamma]S_{ab}[\gamma] - S[\gamma]^2 \big) \quad \text{on $\Sigma$}\, .
\end{equation}
Therefore, in the leading order as $r \to \infty$, the deformation flow equation reduces to
\begin{equation}
   \begin{split}
      \hspace{-.2 cm}  r \frac{\mathrm{d}}{\mathrm{d}r}  {S}^{\text{\tiny{Ren}}}_{\text{bdy}}\Big|_{\Sigma(r)} = & \frac{1}{2 L\kappa_5} \int_{\Sigma(r)} \sqrt{-\gamma} \big(S^{ab}[\gamma]S_{ab}[\gamma] - S[\gamma]^2 \big) \\
        &+ \mathcal{O}(r^{-4})\, .
   \end{split}
\end{equation}
To solve this equation, {we recall $S^{ab}[\gamma]S_{ab}[\gamma] - S[\gamma]^2=\frac{L^4}{4}(R_{ab}[\gamma]R^{ab}[\gamma]-\frac13 R^2[\gamma])$ and \eqref{W2-GB-critical}}, and again use the solution ansatz \eqref{soln-ansatz}. We find the following RG equations
\begin{equation}
    \begin{split}
        & \frac{\d{}}{\d{}r} \left( \frac{\Lambda_4}{\kappa_4} \right) = 0\, , \quad \frac{\d{}}{\d{}r} \left(\frac{1}{\kappa_4}\right) = \frac{L}{3} \left(\frac{L}{r}\right)^3  \frac{\Lambda_4}{\kappa_4}\, , \\
        & \frac{\d{}}{\d{}r} \beta = \frac{L^3}{8 \kappa_5 r} + \frac{L}{2} \left(\frac{L}{r}\right)^3\frac{1}{\kappa_4}\, .
    \end{split}
\end{equation}
We can now integrate these equations 
\begin{equation}
   \begin{split}
       &  {\Lambda_4}=0\, , \quad \kappa_4 = \mathrm{const}\, , \\
       & \beta= \frac{L^3}{8 \kappa_5} \ln\left(\frac{r}{r_0}\right) - \frac{L^2}{4\kappa_4} \left(\frac{L}{r} \right)^2\, .
   \end{split}
\end{equation}
That is, 
\begin{equation}\label{renormalized-induced-gravity}
   \begin{split}
       &{S}^{\text{\tiny{Ren}}}_{\text{bdy}}\Big|_{\Sigma(r)} =  {S}^{\text{\tiny{Ren}}}_{\text{bdy}}\Big|_{\Sigma_0}+\frac{1}{2\kappa_4}  \int_{\Sigma(r)} \sqrt{-\gamma} R[\gamma] \\
      &+\frac{1}{2}\beta(r) \int_{\Sigma(r)} \sqrt{-\gamma}\ W^{abcd}[\gamma] W_{abcd}[\gamma] 
        + \mathcal{O}(r^{-4})\, .
   \end{split}
\end{equation}
It is instructive to compare the above with large $r$ unrenormalized case \eqref{soln-ansatz}: 
\begin{enumerate}
    \item The cosmological constant part, which is divergent in the unrenormalized case, is absent in the renormalized case, i.e., the zero-point energy does not gravitate in the renormalized case. The fact that in the holographically renormalized induced gravity case the \textit{cosmological constant is renormalized to zero}, may point to the mechanism through which the cosmological  RG induced gravity resolves the cosmological constant problem. 
    \item  The gravitational coupling $\kappa_4$ is a scale-independent constant in the renormalized case, whereas it grows quadratically with the scale in the unrenormalized case. This is how the RG-induced gravity setting addresses the non-renormalizability issue of gravity. 
    \item While in both cases the leading part of the coefficient of {Weyl-squared (conformal gravity)} terms has the logarithmic scale dependence, in the renormalized case, there is a subleading piece that falls off quadratically with scale. 

\end{enumerate} 
We close this part with the comment that the induced $4d$ gravity theory has two kinds of subleading terms: (I) Higher order terms in curvature and/or derivatives of curvature; (II) Terms involving combinations of curvature and the EMT tensor ${\cal T}^{ab}$. 
\subsection{Renormalized boundary condition flow}
To complete the above analysis, we should also discuss how the boundary conditions evolve in the renormalized case as we move from the asymptotic boundary to a surface at a large but finite radius. In this setting, the renormalized-rescaled phase space variables are $\gamma_{ab}$ and $\hat{\mathcal{T}}_{ab}$. We begin with Dirichlet boundary conditions at the asymptotic AdS boundary, $\delta \gamma_{ab}\big|_{\Sigma} = 0$, and study the radial evolution of the phase space variables. 

The asymptotic expansion of $\gamma_{ab}$ is given in \eqref{gamma-epsilon} and the asymptotic expansion of $\hat{\mathcal{T}}_{ab} = \epsilon^{-1} \mathcal{T}_{ab}$, which leads to the following variation
\begin{equation}
   \delta \hat{\mathcal{T}}_{ab}(\epsilon) = L^{-4} \delta T_{ab}^{\infty} + \mathcal{O}(\epsilon)\, .
\end{equation}
(Note that the rBY-EMT \eqref{r-BY-EMT} contains a $3h_{ab}/L\kappa_5$ term, which recalling \eqref{gamma-epsilon}, contributes to $\delta \hat{\mathcal{T}}_{ab}(\epsilon)$ by a factor of $-3/2$, changing the $5/2$ in \eqref{delta-gamma-epsilon-1} to +1 in the above.) Finally, combining this result with the first equation in \eqref{delta-gamma-epsilon-1}, we obtain
\begin{equation}
    \delta \gamma_{ab}(\epsilon) =  - \frac{\epsilon^2}{2} \frac{\kappa_{5}}{L^3} \delta (L^4 \hat{\mathcal{{T}}}_{ab}(\epsilon)) +\mathcal{O}(\epsilon^3)\, .
\end{equation}
Expressed in terms of the original (unscaled) variables, this becomes
\begin{equation}\label{brdy-condn-RG-flow-renorm}
  \inbox{  \delta h_{ab}(\epsilon) =  - \frac{1}{2} \frac{\kappa_{5}}{L^3} \delta (L^4 {\mathcal{{T}}}_{ab}(\epsilon)) +\mathcal{O}(\epsilon^2)\, . }
\end{equation}
The difference between the renormalized \eqref{brdy-condn-RG-flow-renorm} and the unrenormalized case \eqref{brdy-condn-RG-flow} lies in their numerical prefactors, $-1/2$ vs. $-1/5$. 
		\bibliographystyle{fullsort.bst}

\bibliography{reference}
\end{document}